\documentclass[reprint,superscriptaddress,amsmath,amssymb,prl,showpacs,floatfix]{revtex4-1}
\usepackage{}
\usepackage{cases}
\usepackage{amsmath}
\usepackage{amssymb}
\usepackage{amsfonts}
\usepackage{amssymb}
\usepackage{dcolumn}
\usepackage{bm}
\usepackage{hyperref}
\usepackage{graphicx}
\usepackage{upgreek}
\usepackage{subfigure}
\usepackage{color}

\begin{document}

\title{Experimental Adiabatic Quantum Factorization under Ambient Conditions Based on a Solid-State Single Spin System}

\author{Kebiao Xu}
\affiliation{CAS Key Laboratory of Microscale Magnetic Resonance and Department of Modern Physics, University of Science and Technology of China, Hefei 230026, China}
\author{Tianyu Xie}
\affiliation{CAS Key Laboratory of Microscale Magnetic Resonance and Department of Modern Physics, University of Science and Technology of China, Hefei 230026, China}
\author{Zhaokai Li}
\affiliation{CAS Key Laboratory of Microscale Magnetic Resonance and Department of Modern Physics, University of Science and Technology of China, Hefei 230026, China}
\affiliation{Synergetic Innovation Center of Quantum Information and Quantum Physics, University of Science
and Technology of China, Hefei, 230026, China}
\author{Xiangkun Xu}
\affiliation{CAS Key Laboratory of Microscale Magnetic Resonance and Department of Modern Physics, University of Science and Technology of China, Hefei 230026, China}
\author{Mengqi Wang}
\affiliation{CAS Key Laboratory of Microscale Magnetic Resonance and Department of Modern Physics, University of Science and Technology of China, Hefei 230026, China}
\author{Xiangyu Ye}
\affiliation{CAS Key Laboratory of Microscale Magnetic Resonance and Department of Modern Physics, University of Science and Technology of China, Hefei 230026, China}
\author{Fei Kong}
\affiliation{CAS Key Laboratory of Microscale Magnetic Resonance and Department of Modern Physics, University of Science and Technology of China, Hefei 230026, China}
\author{Jianpei Geng}
\affiliation{CAS Key Laboratory of Microscale Magnetic Resonance and Department of Modern Physics, University of Science and Technology of China, Hefei 230026, China}
\author{Changkui Duan}
\affiliation{CAS Key Laboratory of Microscale Magnetic Resonance and Department of Modern Physics, University of Science and Technology of China, Hefei 230026, China}
\author{Fazhan Shi}
\affiliation{CAS Key Laboratory of Microscale Magnetic Resonance and Department of Modern Physics, University of Science and Technology of China, Hefei 230026, China}
\affiliation{Synergetic Innovation Center of Quantum Information and Quantum Physics, University of Science
and Technology of China, Hefei, 230026, China}
\author{Jiangfeng Du}
\altaffiliation{djf@ustc.edu.cn}
\affiliation{CAS Key Laboratory of Microscale Magnetic Resonance and Department of Modern Physics, University of Science and Technology of China, Hefei 230026, China}
\affiliation{Synergetic Innovation Center of Quantum Information and Quantum Physics, University of Science
and Technology of China, Hefei, 230026, China}

\begin{abstract}
The adiabatic quantum computation is a universal and robust method of quantum computing. In this architecture, the problem can be solved by adiabatically evolving the quantum processor from the ground state of a simple initial Hamiltonian to that of a final one, which encodes the solution of the problem. By far, there is no experimental realization of adiabatic quantum computation on a single solid spin system under ambient conditions, which has been proved to be a compatible candidate for scalable quantum computation. In this letter, we report on the first experimental realization of an adiabatic quantum algorithm on a single solid spin system under ambient conditions. All elements of adiabatic quantum computation, including initial state preparation, adiabatic evolution, and final state readout, are realized experimentally. As an example, we factored 35 into its prime factors 5 and 7 on our adiabatic quantum processor.
\end{abstract}
\maketitle

The nitrogen-vacancy (NV) center in diamond is an excellent quantum processor and quantum sensor at room temperature \cite{doherty_nitrogen-vacancy_2013}.The spin qubits of NV center are promising for quantum information processing due to fast resonant spin manipulation \cite{fuchs_gigahertz_2009}, long coherence time \cite{balasubramanian_ultralong_2009,Maurer1283}, easy initialization and read-out by laser illumination\cite{gruber_scanning_1997}. Many quantum gates \cite{jelezko_observation_2004,van_der_sar_decoherence-protected_2012,PhysRevLett.112.050503,rong_experimental_2015}, quantum algorithms \cite{PhysRevLett.105.040504}, quantum error corrections \cite{waldherr_quantum_2014,taminiau_universal_2014} and quantum simulations \cite{wang_quantum_2015,PhysRevLett.117.060503} have been demonstrated on it. However, so far no adiabatic quantum algorithm has been realized on this system.

In circuit-model quantum computation, the computational process is implemented by a sequence of quantum gates. In 2000, Farhi {\em et~al}. \cite{farhi_quantum_2000} developed another architecture of quantum computation, i.e., the adiabatic quantum computing (AQC), in which the computational process can be realized through the adiabatic evolution of a system'’s Hamiltonian, and it is proved to be equivalent to circuit model quantum computing \cite{Mizel_2007}.

In contrast to multiplying of large prime numbers, up to now, no efficient classical algorithm for the factorization of large number is known \cite{Knuth:1997:ACP:270146}. Previously, many experimental work on large number factorization have been done based on Shor's algorithm \cite{vandersypen_experimental_2001,lu_demonstration_2007,PhysRevLett.99.250505,politi_shors_2009,martin-lopez_experimental_2012,lucero_computing_2012,monz_realization_2016}.  To demonstrate the AQC on the room temperature single spin system, we take 35 as an example and factored it on the adiabatic quantum processor. The core idea used here is to transform a factorization problem to an optimization problem and solve it under the AQC framework \cite{xu_quantum_2012,PhysRevLett.101.220405}.

Generally, to solve a problem under the AQC framework, first we need to find a problem Hamiltonian $H_p$, and the solution of the problem is encoded in the ground state of $H_p$. We start from the ground state of $H_0$ and the Hamiltonian of the evolution progress is
\begin{equation}\label{eq:AdiaHam1}
\begin{split}
H(t) = &(1-s(t))H_0+s(t)H_p,\\
&s(0)=0, s(T)=1.
\end{split}
\end{equation}
Where $T$ is the total evolution time, and the whole system is governed by the Schr$\rm\ddot{o}$dinger equation
\begin{equation}\label{eq:schrodinger}
i\frac{d}{dt}|{\psi(t)}\rangle=H(t)|{\psi(t)}\rangle.
\end{equation}
According to the adiabatic theorem, if the system evolves slowly enough, i.e., $T\gg 1/g_{min}^2$, where $g_{min}^2$ is the minimum spectral gap of $H(t)$,  it tends to stay on the ground state of $H(t)$. When it reaches $t=T$, the system is on the ground state of $H(T)$ and gives the solution of the problem.
Adiabatic quantum factorization algorithm was first proposed by Burges {\em et~al}. \cite{burges_factoring_2002}, and we adopted an improved version implemented by Xu {\em et~al} \cite{xu_quantum_2012}. Now we explain the method for the factorization of 35.
\renewcommand\arraystretch{1.5}
\begin{table} 
\centering
\begin{tabular}{p{2cm}p{1cm}p{1cm}p{1cm}p{1cm}p{1cm}p{1cm}}
  \hline
    & $2^5$ & $2^4 $ & $2^3 $& $2^2 $& $2^1$ & $2^0 $ \\
  $x$ &   &   &   & 1 & $p$ & 1 \\
  $y$ &   &   &   & 1 & $q$ & 1 \\
  \hline
  &  &  &  & 1 & $p$ & 1 \\
  &  &  &  $q$& $pq$ & $q$ &  \\
    &  & 1 & $p$ & 1 &  &   \\
  \hline
  carries & $z_{45}$ & $z_{34}$  & $z_{23}$ & $z_{12}$ &   &   \\
    &  $z_{35}$  & $z_{24}$  & &  &   &   \\
  \hline
   $x\times y=35$ & 1 & 0 & 0 & 0 & 1 & 1 \\
  \hline
\end{tabular}
\caption{Multiplication table for $5\times 7 = 35$ in binary. The top row represents the significance of each bit. $x$ and $y$ are the multipliers. $z_{ij}$ is the carry bit from the $i$th bit to the $j$th bit, and the row in the bottom is the number we want to factor, i.e., 35.}
\label{binarytable}
\end{table}

Beginning from the multiplication table (see Table.~\ref{binarytable}), we can construct an equation set. In order to solve these equations using fewer qubits, we simplify them further by utilizing some logical constraints. Details about the simplification process can be found in the Supplemental Material \cite{SOM} and finally we can get
\begin{equation}\label{eq:binary6}
p+q=1,
\end{equation}
\renewcommand\arraystretch{1.5}

\noindent and the simplification does't increase the complexity of the quantum algorithm. Until now we can realize the algorithm with two qubits.
According to Burges {\em et~al}. \cite{burges_factoring_2002} the problem Hamiltonian is
\begin{equation}\label{eq:Hp1}
H_p=(\frac{I-\sigma_{1,z}}{2}+\frac{I-\sigma_{2,z}}{2}-I)^{2}=(S_z+I_z)^{2}.
\end{equation}

\noindent We replace $\sigma_{1,z}$ and $\sigma_{2,z}$ with $S_z$ and $I_z$, the electron and nuclear spin operator respectively. Then we drop the identity operator and scale it with a constant $g_1$
\begin{equation}\label{eq:Hp2}
H_p=g_1(2S_zI_z).
\end{equation}
The initial Hamiltonian $H_0$ should be noncommutative with $H_p$, otherwise there will be energy level crossings. Customarily we pick $H_0$ as
\begin{equation}\label{eq:H01}
H_0=g_2(S_x+I_x).
\end{equation}

\begin{figure}
\centering
\includegraphics[width=1\columnwidth]{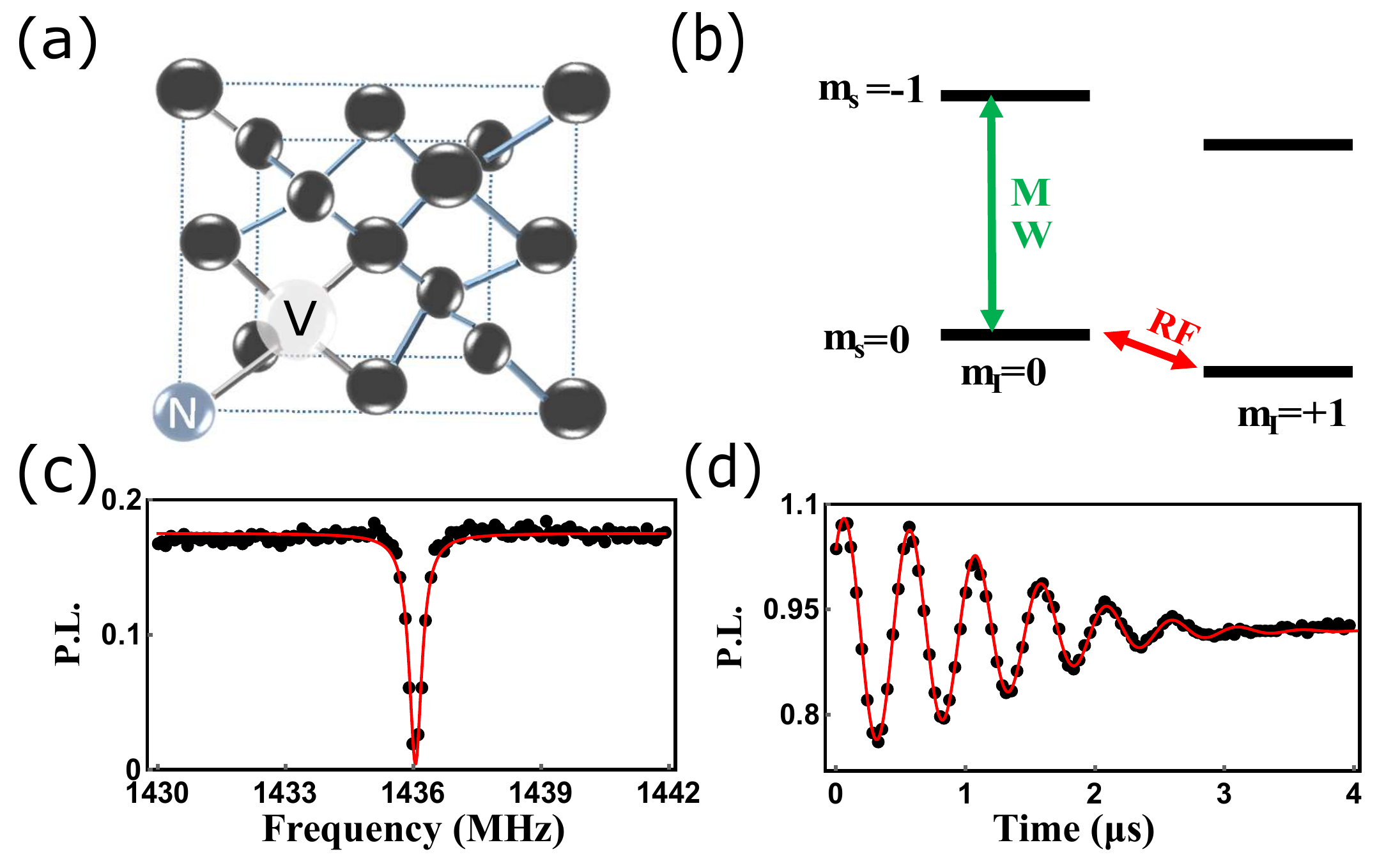}
\caption{(color online). (a) Atomic structure of NV center. (b)Energy level diagram of the subspace that we utilized in the experiment . (c) Optically detected magnetic resonance (ODMR) spectrum of the system, from which we can confirm that the $^{14}$N nuclear spin is polarized. (d) Free induction decay (FID) measurement of the NV center shows a dephasing time $T_2^* = 1.7 \mu s$.
 }\label{system}
\end{figure}

Now we turn to the experimental realization of the factorization algorithm. As we have mentioned, the NV center is a promising candidate for quantum information processing under ambient conditions due to its excellent properties. The experiment was carried out in Type IIa bulk diamond samples with nitrogen impurity concentration $< 5$ ppb and $^{13}$C isotope of natural abundance ($1.1\%$). The atomic structure of the NV center is shown in Fig.~\ref{system}(a). With the direction of the external magnetic field aligned along the symmetry axis, the Hamiltonian of the NV center electron spin coupled with the $^{14}$N nuclear spin is
\begin{equation}\label{eq:Hamiltionian1}
H = D{S_z}^2+\gamma_eB_zS_z+Q{I_z}^2+\gamma_nB_zI_z+\hat{S}\tilde{A}\hat{I}.
\end{equation}
Where $D=2.87$ GHz is the zero field splitting and $Q=-4.95$ MHz is the nuclear quadrupolar splitting. $\gamma_e$ and $\gamma_n$ are electron and nuclear gyromagnetic ratio respectively, and $\tilde{A}$ is the hyperfine interaction tensor between the electron spin and the nuclear spin. Fig.~\ref{system}(b) depicts part of the energy level diagram of the system. We picked 4 out of 9 levels to perform the experiment and encoded them as two qubits. We use ``0" and ``1" to represent the two quantum states of the NV center and the nuclear spin. For example, $|01\rangle$ denotes the  $|{m_s=0,m_I=0}\rangle$ state.

\begin{figure}
\centering
\includegraphics[width=1\columnwidth]{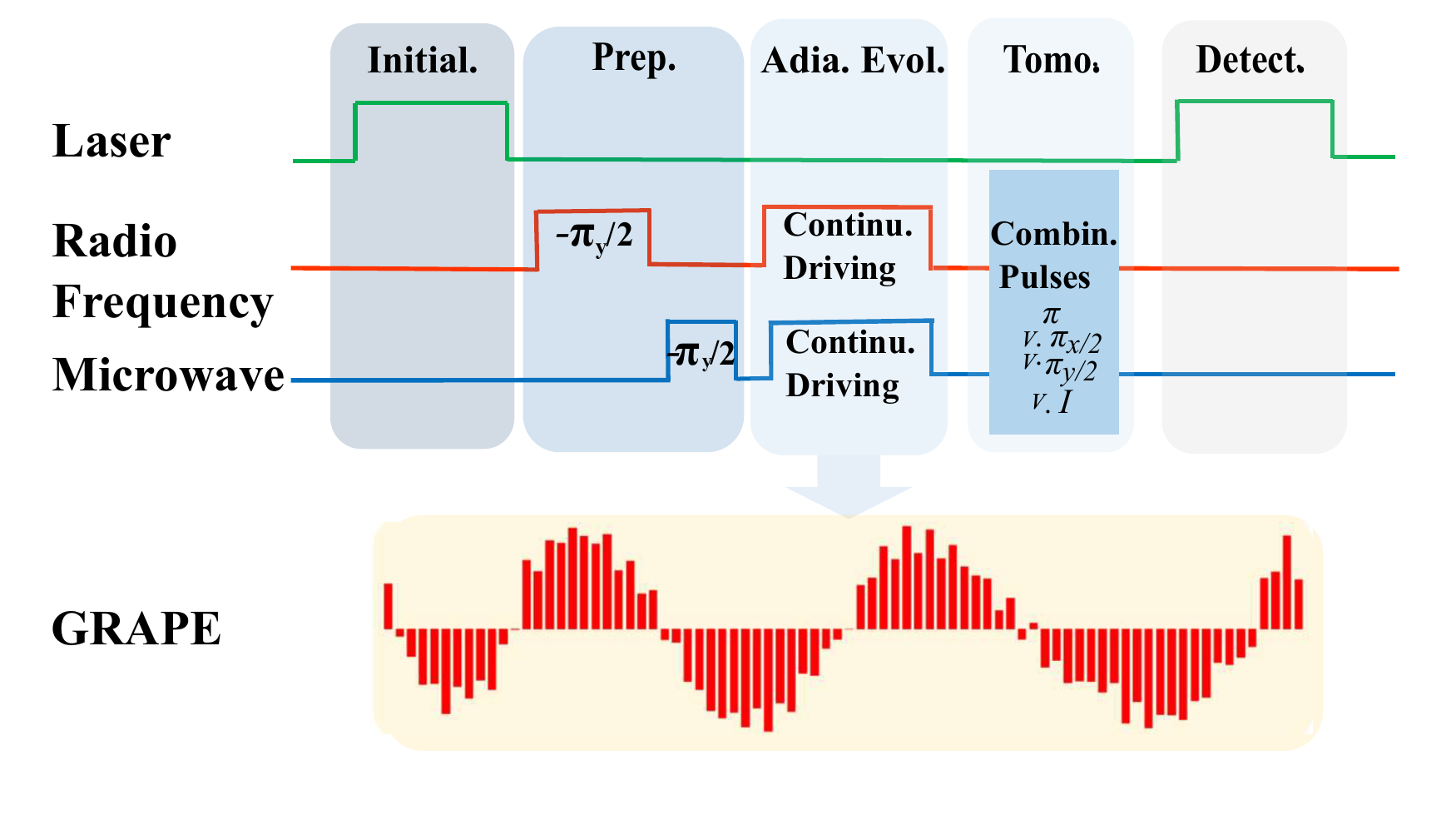}
\caption{(color online). The pulse sequence of the adiabatic factorization process. Under the condition that the external magnetic field is 510 G aligned with the NV axis, the  532 nm laser pulse initialize both the electron spin and the $^{14}$N nuclear spin. To prepare the initial state to $\Psi_{i} = \frac{1}{2}(|{0}\rangle-|{1}\rangle)(|{0}\rangle-|{1}\rangle)$, RF and MW pulses are applied successively. The above panel shows that the adiabatic evolution can be realized through continuously applying the RF and MW pulses. In the actual experiment, evolution of the state is driven by the optimal control pulse instead. In the lower panel is the optimal control pulse applied on nuclear spin.
 }\label{Pulsesequence}
\end{figure}

In order to construct the adiabatic evolution Hamiltonian, a similar quantum simulation approach proposed in Ref \cite{PhysRevLett.117.060503} can be adopted here. By simultaneously applying RF and MW driving, the system Hamiltonian was transformed to \cite{ref_fei}
\begin{equation}\label{eq:HNV_rot_1}
\begin{aligned}
H_{NV}^{rot}=(\delta_{MW}+\delta_{RF}){\hat{\mathbb{I}}}+2\Omega_{MW}S_xI_z\\
-\delta_{RF}I_{z}+\Omega_{RF}I_x-\delta_{MW}S_z.
\end{aligned}
\end{equation}

\noindent In which $\Omega_{\rm MW}$ and $\Omega_{\rm RF}$ are the Rabi frequencies, $\delta_{\rm RF}$ and $\delta_{\rm MW}$ are the detunings, if we set $\delta_{\rm RF}=0$, $2\Omega_{\rm MW}\equiv 2g_1$, $\Omega_{\rm RF}=-\delta_{\rm MW}\equiv g_2$ and drop the identity operator, then
\begin{equation}\label{eq:HNV_rot_2}
H_{NV}^{rot}=2g_1S_xI_z+g_2(I_x+S_z).
\end{equation}
Next a Hadamard gate is applied on the electron spin subspace, which changes the basis form $|{0}\rangle, |{1}\rangle$ to $|{0}\rangle+|{1}\rangle, |{0}\rangle-|{1}\rangle$, and exchanges $S_x$ and $S_z$, then we get
\begin{equation}\label{eq:HNV_rot_3}
H_{NV}^{rot}=2g_1S_zI_z+g_2(I_x+S_x).
\end{equation}
This exactly recovers the Eq.\eqref{eq:Hp2} and Eq.\eqref{eq:H01}.
To overcome the limitation of coherence time, the optimal control is adopted here to replace the adiabatic evolution part, which is robust to noise and guarantees high fidelity. Recently, the shaped pulse technique, which is used for optimal controls in our experiments, has been realized by serval NV-based quantum computation proposals \cite{rong_experimental_2015,waldherr_quantum_2014,dolde_high-fidelity_2014}.

The experiment consists of four stages. As the external magnetic field is 510 G, we can polarize the NV electron spin and $^{14}$N nuclear spin simultaneously by applying 532 nm laser \cite{PhysRevLett.102.057403}. Then we applied a $-{\pi_y}/{2}$  microwave (MW) and radio-frequency (RF) pulse to prepare the initial state $\Psi_{i} = \frac{1}{2}(|{0}\rangle-|{1}\rangle)(|{0}\rangle-|{1}\rangle)$. Where $-{\pi_y}/{2}$ is a ${\pi}/{2}$ rotation around the -y axis. Next the adiabatic process was approximated by shaped pulses. As the gyromagnetic ratio of the electron spin is three orders of magnitude larger than that of the nuclear spin, previous optimal control pulses were applied on electron and nuclear spin separately. Here we not only applied shaped pulses to realize optimal control over the hybrid spin system, but also first time applied shaped pulses on electron and nuclear spins simultaneously. Detailed description of the shaped pulses is in the Supplementary Material \cite{SOM}. The solution of the equation can be extracted from measuring the population distribution of the final state. But in fact we carried out state tomography and measured all terms in the density matrix to check the consistency with theory. Totally $16$ combinations of MW and RF $\{\pi,\frac{\pi}{2}_x,\frac{\pi}{2}_y,\hat{\mathbb{I}}\}$ pulses are applied to readout the diagonal and off-diagonal elements of the final state density matrix.

\begin{figure}
\centering
\includegraphics[width=1\columnwidth]{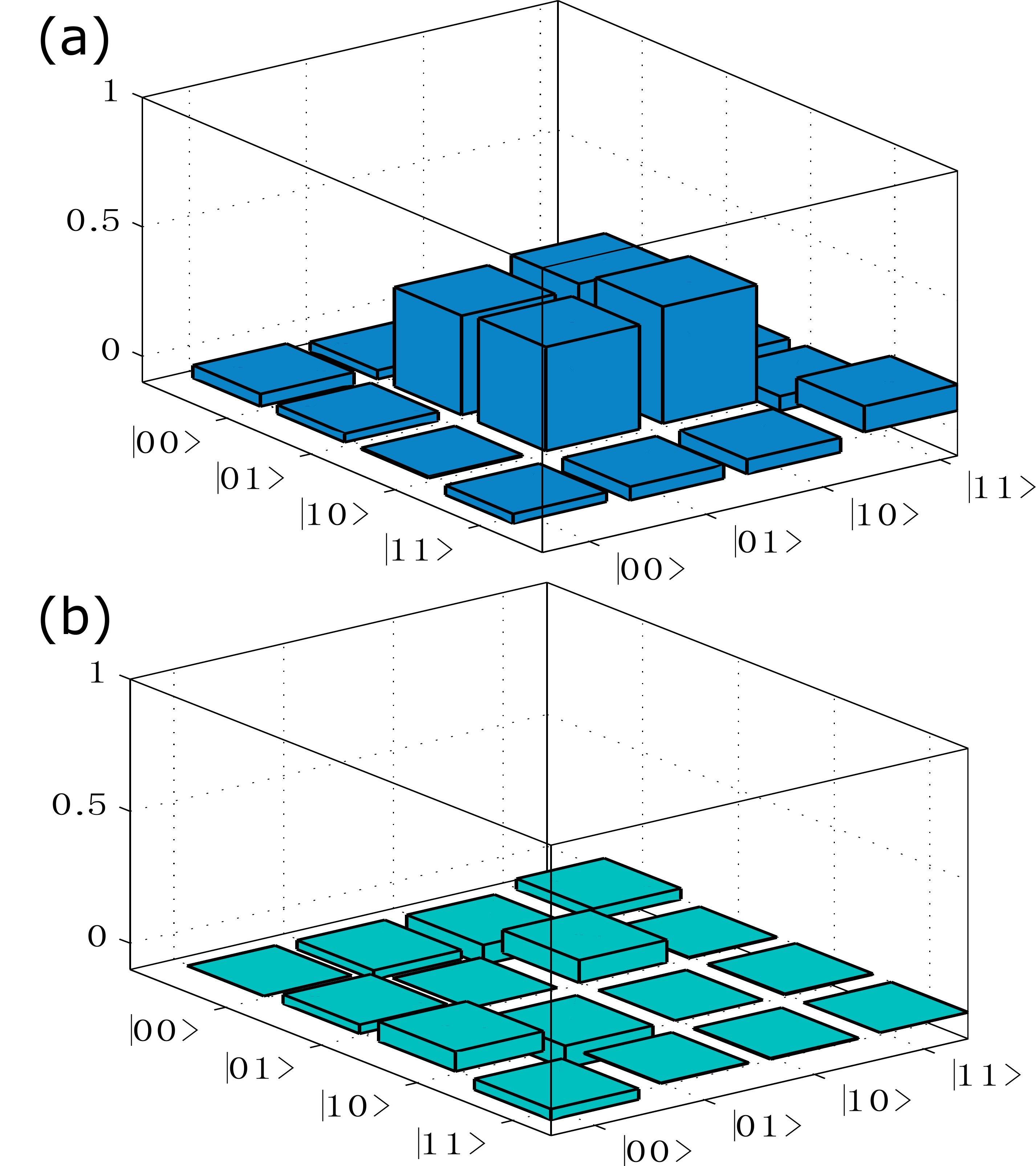}
\caption{(color online). Final state density matrix.
(a) The real and (b) the imaginary part of the density matrix. From this final state we can infer the answer of the factorization is \{\textit{p}=1,\ \textit{q}=0\} or \{\textit{p}=0,\ \textit{q}=1\}.
}\label{finalstate}
\end{figure}

\begin{figure}
\centering
\includegraphics[width=1\columnwidth]{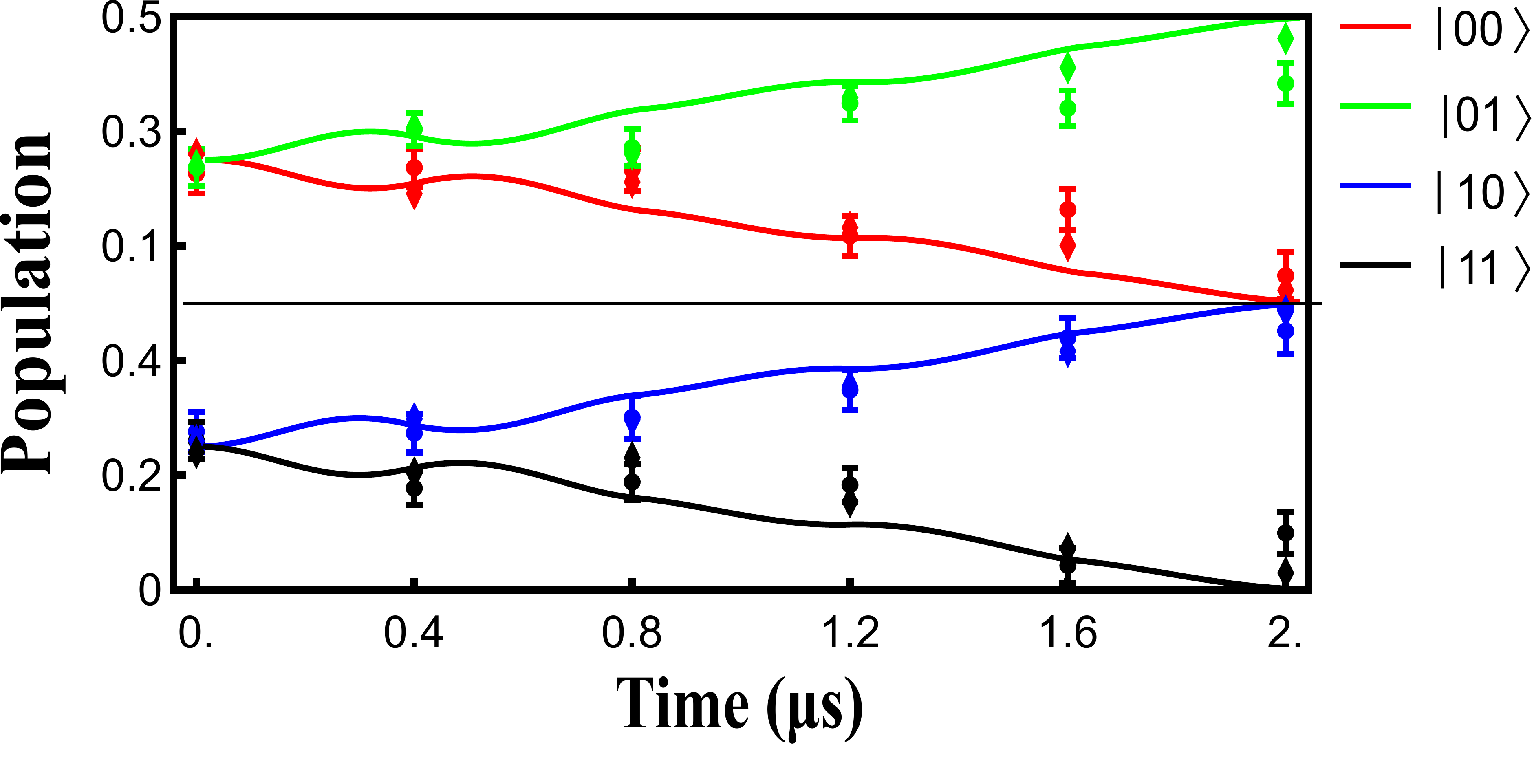}
\caption{(color online). The populations on computational basis of the system during the process of the adiabatic factorization. The lines are theoretical predictions with prefect initialization and control. The diamonds are the theoretical calculations taking imperfect initialization and control into consideration and the dots are got form experimental data.
 }\label{evolution}
\end{figure}

Fig.~\ref{finalstate} shows the real and imaginary part of the final state. The fidelity between the experimental and the ideal final state $\Psi_{f} = \frac{1}{\sqrt{2}}(|{01}\rangle+|{10}\rangle)$ is 0.81(6). To show the adiabaticity of the whole process, we sampled the evolution with 6 measurements of the diagonal elements, as exhibited in Fig.~\ref{evolution}. The lines and dots represent the results of theoretical calculations and experimental measurements respectively. The deviation of the dots from the lines at some points is due to the imperfect initialization and control pulses. If we take the error of polarization and the amplitude fluctuations of  RF and MW pulses into consideration (as depicted by the diamonds), the experimental data fits well with the calculations. To further confirm that the state during evolution is really on the ground state of the Hamiltonian, we even calculated the energy and the fidelity between the actual state and the ground state of the Hamiltonian during the evolution process (Fig. S2 in in the Supplemental Material \cite{SOM}). The population of the final state concentrates on $|01\rangle$ and $|10\rangle$, which denotes that the solution of the factorization problem is $ \{p=0,\ q=1\}$ or $\{p=1,\ q=0\}$, i.e., the multipliers are $ \{x=5,\ y=7\}$ or $ \{x=7,\ y=5\}$.

In conclusion, we have presented the experimental demonstration of adiabatic factorization on a solid state single spin system under ambient conditions. We experimentally factored 35  and got the results with high fidelity. To achieve this, we have improved the optimal control technology and enabled it to be used on an electron-nuclear hybrid system, which is considered to be challenging due to the large mismatch of the electron and nuclear spin gyromagnetic ratio. Furthermore, as mentioned by Dattani {\em et~al}. in Ref. \cite{dattani_quantum_2014}, this process factors not only 35, but also a kind of integers, which can be six-digit numbers or larger. Besides, the optimal control technique used here can also be extended to other single-spin systems \cite{koehl_room_2011,pla_high-fidelity_2013,yin_optical_2013}, as they share the similar control method based on spin magnetic resonance. Therefore, we envision that this work could have further implications for the area of precise quantum control and quantum information science.

The authors thank Mingliang Tian and Haifeng Du for their kindly help on the fabrication of solid immersion lens (SIL). This work was supported by the 973 Program (Grant No.~2013CB921800, No.~2016YFA0502400), the National Natural Science Foundation of China (Grants No.~11227901, No.~31470835, No.~11575173 and No.~91636217), the China Postdoctoral Science Foundation, the CAS (Grant No.~XDB01030400, No.~QYZDY-SSW-SLH004), and the Fundamental Research Funds for the Central Universities (WK2340000064).

K. X., T. X. and Z. L. contributed equally to this work.

\end{document}